\documentclass{epl}

\usepackage{graphicx}
\usepackage{dcolumn}
\usepackage{bm}

\title{Spin Josephson effect in ferromagnet/ferromagnet tunnel junctions}
\shorttitle{Spin Josephson effect}
\author{Flavio S. Nogueira \and Karl-Heinz Bennemann}
\institute{Institut f\"ur Theoretische Physik,
Freie Universit\"at Berlin, Arnimallee 14, D-14195 Berlin, Germany}


\begin{document}

\maketitle

\begin{abstract}
We consider the tunnel current between two ferromagnetic metals 
from a perspective similar to the one used in 
superconductor/superconductor tunnel junctions. We use 
fundamental arguments to derive  
a Josephson-like tunnel spin current 
$I_J^{\rm spin}\propto\sin(\theta_1-\theta_2)$. 
Here the phases are associated with the planar contribution to 
the magnetization, 
$\langle c_\uparrow^\dagger c_\downarrow\rangle\sim e^{i\theta}$.  
The crucial step in our analysis is the fact that the 
$z$-component of the spin is canonically conjugate to 
the phase of the planar contribution: $[\theta,S^z]=i$. This 
is the counterpart to the commutation 
relation $[\varphi,N]=i$ in superconductors, where 
$\varphi$ is the phase associated with the superconducting 
order parameter and $N$ is the Cooper pair number operator. We 
briefly discuss the  
experimental consequences of our theoretical analysis.  
\end{abstract}

In recent years the possibility of using the spin degrees of freedom 
in addition to the charge degrees of freedom in electronic devices has 
open new perspectives in condensed matter physics. Indeed, the field 
now known as {\it spintronics} or {\it magnetoelectronics} 
is rapidly developing and there are even commercial devices available 
\cite{Spintronics}. If the current density due to an electron of spin $\sigma$ is 
${\bf j}_\sigma$, then the usual electric current is just the sum 
of the electronic currents for each spin degree of freedom, 
${\bf j}^{\rm charge}=\sum_\sigma {\bf j}_\sigma$. 
The spin current density, on the 
other hand, can be defined by 
${\bf j}^{\rm spin}=\sum_\sigma \sigma{\bf j}_\sigma$. 
Although such a definition of spin current density is used very often, it 
should be kept in mind that that the true spin 
current density is actually a tensor and not simply a vector like 
the charge current density. Indeed,  
if ${\bf M}$ is the magnetization, the continuity equation
for the spin current is given by 
$\partial M_i/\partial t+\partial_\mu J_{i\mu}^{\rm spin}=0$, 
where $J_{i\mu}^{\rm spin}$ is the spin current density tensor. 

The most familiar situation in the theory of electronic transport 
corresponds to the case where a charge current flows in 
the system and the spin current is zero. However, the opposite situation is 
also possible in some ferromagnetic 
systems, i.e., a spin current flowing in the absence of any 
charge current \cite{Hirsch,Brataas,Wang}. A relevant problem 
to study in this context is the behavior of the spin current 
across a tunnel junction between 
two ferromagnetic metals \cite{Sloncz}, especially whether 
there is phase coherence of the spin current across the junction. 
Phase coherent charge transport is a well known feature in tunnel 
junctions between superconductors and is responsible for the 
Josephson effect \cite{Josephson}. In this paper we will study a  
similar effect for a spin current in 
ferromagnet/ferromagnet (FM/FM) tunnel junctions.   
Recently it was pointed out that a {\it charge} Josephson-like effect 
would be possible in a spin valve \cite{Su}, but our 
analysis below does not confirm such a prediction.   
The spin Josephson effect in itself is not new and was observed some 
time ago in $^3$He-B weak links \cite{Borovik,Fomin}. The 
observation of this effect establishes the existence of 
spin supercurrents in $^3$He-B. It is plausible that 
a similar effect may occur in other systems, like in tunnel junctions 
involving ferromagnetic metals. Another possibility is to use thin film 
helimagnets, where spin transport 
without dissipation was discussed recently \cite{Konig}. 

The main result of this letter will be a {\it microscopic}  
derivation of the spin Josephson effect in FM/FM tunnel 
junctions. The key observation in our analysis is 
that in a ferromagnetic metal spin flip averages 
like $\langle c_\uparrow^\dagger c_\downarrow\rangle$ are nonzero 
and play a role analogous to the 
superconducting order parameter 
$\langle c_\uparrow^\dagger c_\downarrow^\dagger\rangle$.  
This same analogy was used recently in discussions of a   
Josephson-like effect 
\cite{Wen,Ezawa,Stern,Balents} observed 
in double layer quantum Hall pseudo-ferromagnet \cite{Spielman}. 
There the spin label is associated with the layer such that an 
analogy with exciton superfluids \cite{Lozovik}  
is also possible \cite{Fogler}.    
 
In order to make our analysis of 
the spin Josephson effect self-contained, we first recall 
some basic facts about superconductor/superconductor (SC/SC) tunnel 
junctions. There the phase plays a crucial role leading to the 
celebrated Josephson effect 
\cite{Anderson}. Indeed, the Josephson effect is a consequence of the 
fact that the phase $\varphi$
of the order parameter is canonically 
conjugate to the Cooper pair number operator $N$, 
$[\varphi,N]=i$,
which implies the uncertainty relation $\Delta\varphi\Delta N\sim 1$ 
\cite{Anderson}. The superconducting current is given by 
$I_{\rm SC}=2e\langle(\dot{N}_2-\dot{N}_1)\rangle$, where $e$ is the electric 
charge and $N_1$ and $N_2$ are the Cooper pair number operators 
for the superconductors in each side of the tunnel junction. 
Using then the Hamiltonian 
for the tunnel junction  

\begin{equation}
\label{H-sc}
H=-E_J\cos(\varphi_1-\varphi_2)+\frac{2e^2}{C}(N_1-N_2)^2,
\end{equation}
where $C$ is the capacitance of the tunnel junction, and treating 
$N_i$ and $\varphi_i$ as classical canonically conjugate variables, 
we easily obtain $I_J^{\rm SC}=(E_J/e)\sin\Delta\varphi$,  
$\Delta\dot{\varphi}=2eV$, where $\Delta\varphi\equiv\varphi_1-\varphi_2$ and 
$V=2e(N_1-N_2)/C$ \cite{Note1}.     

In the case of FM/FM tunnel junctions it is the $z$-component of 
the spin that is canonically conjugated to the phase: 
$[\theta,S^z]=i$. This follows from writing the raising and lowering  
spin operators as \cite{Villain}

\begin{eqnarray}
\label{repres}
S_{+}&=&\exp(i\theta)~\sqrt{\left(S+\frac{1}{2}\right)^2
-\left(S^z+\frac{1}{2}\right)^2}, 
\nonumber\\
S_{-}&=&\sqrt{\left(S+\frac{1}{2}\right)^2
-\left(S^z+\frac{1}{2}\right)^2}~\exp(-i\theta).
\end{eqnarray}
Thus, we obtain the uncertainty relation $\Delta\theta\Delta S^z\sim 1$.
Furthermore, we can derive the commutation relation 
$[S^z,e^{\pm im\theta}]=\pm e^{\pm im\theta}$ which in turn 
implies the commutation relations defining the algebra of 
spin operators. Using the representation (\ref{repres}) we 
can establish a classical Hamiltonian for the FM/FM tunnel 
junction out of which 
a spin current  
$I_J^{\rm spin}=\mu_B\langle\dot{S}_R^z-\dot{S}_L^z\rangle$ 
is derived: 

\begin{equation}
\label{H-fm}
H=-E_JS^2\cos(\theta_L-\theta_R)
+\frac{\mu_B^2}{2C_s}(S_L^z-S_R^z)^2,
\end{equation}
where for convenience we have introduced the ``spin capacitance'' $C_s$, 
which will be specified later in the discussion of the microscopic 
origin of the effect. The coeffcient $E_JS^2$ of the 
first term of Eq. (\ref{H-fm}) corresponds in fact to an 
exchange energy between the magnetizations in both sides of the 
tunnel junction. The value of $E_JS^2$ will be determined later 
in terms of parameters of the microscopic model.
   
Using the classical Hamiltonian equations of motion we easily 
derive: 

\begin{equation}
\label{spinjos}
\Delta\dot{\theta}=2\mu_B V_s, 
~~~~I_J^{\rm spin}=(2E_JS^2/\mu_B)\sin\Delta\theta,  
\end{equation}
where we have defined the spin 
voltage as $V_s=(\mu_B/C_s)(S_L^z-S_R^z)$.  
Thus, a spin current in a FM/FM tunnel junction 
would behave in the same way as 
the superconducting current in the SC/SC tunnel junction, see Table I.  

\begin{table}
\caption{Analogies between SC/SC and FM/FM tunnel junctions}
\begin{center}
\begin{tabular}{cc}
SC/SC junction & FM/FM junction \\
\hline
$[\varphi,N]=i$ & $[\theta,S_z]=i$ \\
Voltage $V$ & Spin voltage $V_s$ \\
$\Delta\dot{\varphi}=2eV$ & $\Delta\dot{\theta}=2\mu_B V_s$ \\
$I_J^{\rm SC}=(E_J/e)\sin\Delta\varphi$ & $I_J^{\rm spin}=(2E_JS^2/\mu_B)\sin\Delta\theta$
\end{tabular}
\end{center}
\end{table}

Now that we have motivated the problem on the basis of a  
straightforward physical analogy, 
we are ready to derive microscopically the spin Josephson 
effect in FM/FM tunnel junctions. 
Before starting, it is worth to  
compare once more the role of the phases in  
superconductors and ferromagnets. In 
superconductors a phase transformation of the 
electron destruction operators involves the particle number 
operator:

\begin{eqnarray}
\label{transf-sc}
c_\sigma(\varphi)&=&\exp\left[i\frac{\varphi}{2}(n-1)\right]c_\sigma(0)
\exp\left[-i\frac{\varphi}{2}(n-1)\right]\nonumber\\
&=&e^{i\varphi/2}~c_\sigma(0),
\end{eqnarray}
where $n=\sum_\sigma c_\sigma^\dagger c_\sigma$ 
and from the first to the second line we have just 
solved the equation of motion 
$\partial c_\sigma/\partial\varphi=i[(n-1)/2,c_\sigma]$. 
Similarly, for a ferromagnetic 
metal the relevant transformation is

\begin{eqnarray}
\label{transf-fm}
c_\sigma(\theta)&=&\exp\left[i\frac{\theta}{2}(n_\uparrow-n_\downarrow)\right]c_\sigma(0)
\exp\left[-i\frac{\theta}{2}(n_\uparrow-n_\downarrow)\right]\nonumber\\
&=&e^{i\sigma\theta/2}~c_\sigma(0).
\end{eqnarray}
Eq. (\ref{transf-sc}) is thus associated to gauge transformations 
$c_\sigma=e^{i\varphi/2}\tilde{c}_\sigma$ out of which we 
obtain that 
$\langle c_\uparrow^\dagger c_\downarrow^\dagger\rangle=e^{-i\varphi}
\langle\tilde{c}_\uparrow^\dagger\tilde{c}_\downarrow^\dagger\rangle$.  
Eq. (\ref{transf-fm}), on the other hand, yields  
$\langle c_\uparrow^\dagger c_\downarrow\rangle=e^{-i\theta}
\langle\tilde{c}_\uparrow^\dagger\tilde{c}_\downarrow\rangle$.  

We will consider the ferromagnetic metals in each side of 
the junction in the simple framework of the Stoner model 
for itinerant ferromagnetism \cite{Stoner}. 
The interaction is given simply by 

\begin{equation}
\label{interac}
H_I=U\int d^3r ~n_\uparrow({\bf r})n_\downarrow({\bf r}),
\end{equation}
which can be rewritten as

\begin{eqnarray}
\label{HS}
H_I'&=&\int d^3r\left\{\frac{3}{2U}[m^2({\bf r})+|\Delta({\bf r})|^2]
-c_\uparrow^\dagger({\bf r})c_\downarrow({\bf r})\Delta({\bf r})
\right.\nonumber\\
&-&\left.
\Delta^\dagger ({\bf r})c_\downarrow^\dagger({\bf r})c_\uparrow({\bf r})
-m({\bf r})[n_\uparrow({\bf r})-n_\downarrow({\bf r})]\right\},
\end{eqnarray}
upon a Hubbard-Stratonovich transformation. 
It is easy to see from the equations of motion for the auxiliary fields 
that mean-field theory corresponds to 

\begin{equation}
\label{spinop}
\Delta^\dagger=(2U/3)\langle S_{+}\rangle,~~~
\Delta=(2U/3)\langle S_-\rangle,~~~ m=(2U/3)\langle S_z\rangle,
\end{equation} 
where the spin operators 
are defined in terms of fermion operators as  
$S_{+}=c_\uparrow^\dagger c_\downarrow$ and 
$S_z=(n_\uparrow- n_\downarrow)/2$. 

The electron field operators in the left and right subsystems 
will be denoted as $c_\sigma$ and $f_\sigma$, respectively. 
The bosonic auxiliary fields in the left and 
right subsystems will be distinguished respectively by labels $L$ and 
$R$ and assumed to be uniform in a mean-field approximation. 
Futhermore, for simplicity we will also 
assume that the left and right ferromagnetic metals are 
the same. Thus, we will take $m_L^2=m_R^2=m^2$ and 
$|\Delta_L|=|\Delta_R|=|\Delta|$, but in general 
$m_L\neq m_R$, since we can have $m_L=m_0$ and $m_R=-m_0$. 
Also, the phases of 
$\Delta_L$ and $\Delta_R$ are generally not the same. The 
tunnel Hamiltonian is given as usual by 
$H_T=\sum_{{\bf k},{\bf p}}\sum_\sigma(T_{{\bf kp}}c_{{\bf k}\sigma}^
\dagger f_{{\bf p}\sigma}+{\rm h.c.})$, where $T_{{\bf kp}}$ is the  
tunnel matrix element. The charge current is given by 
$I^{\rm charge}=-e\langle\dot{N}_L(t)\rangle$, where 
$N_L=\sum_{{\bf k},\sigma}c_{{\bf k}\sigma}^
\dagger c_{{\bf k}\sigma}$ and 
$\dot{N}_L=i\sum_{{\bf k},{\bf p}}\sum_\sigma(T_{{\bf kp}}c_{{\bf k}\sigma}^
\dagger f_{{\bf p}\sigma}-T_{{\bf kp}}^*f_{{\bf p}\sigma}^\dagger 
c_{{\bf k}\sigma})$, while the spin current is given 
by $I^{\rm spin}=-\mu_B\langle\dot{S}_L^z\rangle$ 
where $S_L^z=\sum_{{\bf k},\sigma}\sigma c_{{\bf k}\sigma}^
\dagger c_{{\bf k}\sigma}$ and 
$\dot{S}_L^z=i\sum_{{\bf k},{\bf p}}\sum_\sigma
\sigma(T_{{\bf kp}}c_{{\bf k}\sigma}^
\dagger f_{{\bf p}\sigma}-T_{{\bf kp}}^*f_{{\bf p}\sigma}^\dagger 
c_{{\bf k}\sigma})$.  

We will evaluate the spin current using linear response 
theory. In this approximation we have 
$I^{\rm spin}=2\mu_B{\rm Im}[\Xi(eV+i\delta)]$, where  
$\delta\to 0+$. In terms of the Matsubara formalism 
we have  

\begin{eqnarray}
\label{Xi}
\Xi(i\omega)&=&\frac{1}{\beta}\sum_{\omega_1}
\sum_{{\bf k},{\bf p}}|T_{{\bf kp}}|^2
\left[\sum_\sigma \sigma G_L^\sigma({\bf k},i\omega_1)
G_R^\sigma({\bf p},i\omega_1-i\omega)
\right.\nonumber\\
&+&\left.F_L^{\downarrow\uparrow}({\bf k},i\omega_1)
F_R^{\uparrow\downarrow}({\bf p},i\omega_1-i\omega)
-F_L^{\uparrow\downarrow}({\bf k},i\omega_1)
F_R^{\downarrow\uparrow}({\bf p},i\omega_1-i\omega)\right].
\end{eqnarray}
The Green functions appearing in Eq. (\ref{Xi}) are given 
in the mean-field approximation by 

\begin{eqnarray}
\label{Gs}
G_{L}^\sigma({\bf k},i\omega)&=&\frac{i\omega-\epsilon_{\bf k} 
+\sigma m_L}{(i\omega-\epsilon_{\bf k})^2-m^2-|\Delta|^2},
\nonumber\\
F_L^{\downarrow\uparrow}({\bf k},i\omega)
&=&\frac{\Delta_L}{(i\omega-\epsilon_{\bf k})^2-m^2-|\Delta|^2},
\nonumber\\
F_L^{\uparrow\downarrow}({\bf k},i\omega)&=&
\frac{\Delta_L^\dagger}{(i\omega-\epsilon_{\bf k})^2-m^2-|\Delta|^2},
\end{eqnarray}
with similar expressions for the right side Green functions. 
The dispersion energies are $\epsilon_{\bf k}=e_{\bf k}-\mu_L$ 
and $\epsilon_{\bf p}=e_{\bf p}-\mu_R$, where 
$\mu_L-\mu_R=eV$. 
The 
spin current can be separated into two parts, one due to 
direct spin current tunneling and another one due 
to spin flip tunneling. It is the latter that is responsible 
for the spin Josephson effect and arises from the 
product between the $F_{L,R}^{\alpha\beta}$ Green functions.  
The direct spin current vanishes at zero voltage while 
the spin flip current does not. Explicit evaluation gives 

\begin{equation}
\label{spin-Jos-cur}
I_J^{\rm spin}=\frac{\pi \mu_B|\Delta|^2|T|^2}{2(m^2+|\Delta|^2)}
S(eV,\sqrt{m^2+|\Delta|^2})\sin\Delta\theta
\end{equation}
where   
$\Delta\theta$ is the phase difference between 
the left and right subsystem as before. The function 
$S(a,b)$ is given by 

\begin{eqnarray} 
\label{ReS2}
S(a,b)&=&\int du\int dv 
~\rho(u)~\rho(v)\left[
\frac{f(v+b)+
f(v-b)-
f(u+b)-
f(u-b)}{a+v-u}
\right.\nonumber\\
&-&\left.\frac{f(v+b)-
f(u-b)}{a+2b+v-u}-
\frac{f(v-b)-
f(u+b)}{a-2b+v-u}\right],
\end{eqnarray}
where $\rho(\epsilon)$ is the density of states and 
$f(\epsilon)$ is the Fermi-Dirac distribution function. 
The function $S(a,b)$ satisfy the properties $S(a,0)=0$ 
and $S(0,b)\neq 0$. 
Eq. (\ref{spin-Jos-cur}) establishes microscopically 
the spin Josephson effect in FM/FM tunnel junctions. 
The Josephson-like spin tunnel current is nonzero 
only if $|\Delta|\neq 0$. Furthermore, it is maximum for 
$m=0$, which corresponds to the case of a tunnel junction between 
planar ferromagnets.

From Eq. (\ref{spin-Jos-cur}) we can easily determine 
the exchange magnetic energy across the junction, since 
$I_J^{\rm spin}=\partial E/\partial\Delta\theta$, where 
$E$ is the energy of the tunnel junction. In this way we 
obtain that the coefficient $E_JS^2$ appearing in Eq. 
(\ref{H-fm}) is given by the coefficient of the 
sine of Eq. (\ref{spin-Jos-cur}) with $V=0$.    

The spin Josephson effect as given by 
Eq. (\ref{spin-Jos-cur})  
implies a spin current even in the absence of voltage. Now we 
will show that when $m_L-m_R$ is nonzero the above effect is 
actually an AC-like effect.  
In order to show this we have to use a gauge transformation based 
on Eq. (\ref{transf-fm}) in the Lagrangians for the left and right 
subsystems, i.e., we set 
$c_{{\bf k}\sigma}=e^{i\sigma\theta_L/2}\tilde{c}_{{\bf k}\sigma}$ and 
$f_{{\bf k}\sigma}=e^{i\sigma\theta_R/2}\tilde{f}_{{\bf k}\sigma}$.  
For instance, we obtain for the left subsystem 

\begin{equation}
\label{LLfm}
L_L^{\rm MF}=\sum_{{\bf k},\sigma}\tilde{c}_{{\bf k}\sigma}^\dagger 
\left(i\partial_t-\frac{\sigma}{2}\partial_t\theta_L+\sigma m_L
+\frac{eV}{2}\right)\tilde{c}_{{\bf k}\sigma}-H_L^{\rm MF},
\end{equation} 
with a similar expression for the right subsystem. In analogy with 
the Josephson effect in superconductors, we can compensate 
for the phase transformation by writing 
$\partial_t\theta_L=2m_L$ and $\partial_t\theta_R=2m_R$, out of 
which we obtain 

\begin{equation}
\label{ACeffect}
\Delta\dot{\theta}=2(m_L-m_R)=2\mu_B V_s,
\end{equation}
where we have defined a ``spin voltage'' $V_s\equiv(m_L-m_R)/\mu_B$.   
Using the expression for $m$ in Eq. (\ref{spinop}) for the left and 
right systems and comparing Eq. (\ref{ACeffect} with Eq. (\ref{spinjos}), 
we obtain that the ``spin capacitance'' is given by

\begin{equation}
\label{spincap}
C_s=\frac{3}{2}\frac{\mu_B^2}{U}.
\end{equation}  
It is worth to discuss once more 
the similarities between FM/FM and SC/SC tunnel junctions.  
In the latter the time 
evolution of the phase difference is proportional to the 
average particle number difference \cite{Tinkham}, while in the former it is 
proportional to the difference of the $z$-projection of the 
magnetization across the junction. Note that in superconductors a 
Hubbard-Stratonovich transformation similar to the one in Eq. (\ref{HS}) 
holds with the spin operators replaced by the pseudo-spin operators 
$\tilde{S}_+=c_\uparrow^\dagger c_\downarrow^\dagger$, 
$\tilde{S}_-=\tilde{S}_+^\dagger$, 
$\tilde{S}_z=(n_\uparrow+n_\downarrow-1)/2$.
    
If $m_L^2=m_R^2=m^2$, the AC-like spin Josephson 
effect occurs only if the left and right subsystems have 
the spin projection on the $z$-axis anti-parallel to one another.     
By solving Eq. (\ref{ACeffect}) with 
$m_L=-m_R=m$ and using Eq. (\ref{spin-Jos-cur}) 
we obtain the important result 

\begin{equation}
\label{AC-cur}
I_J^{\rm spin}=\frac{\pi \mu_B|\Delta|^2|T|^2}{2(m^2+|\Delta|^2)}
S(eV,\sqrt{m^2+|\Delta|^2})\sin(\Delta\theta_0+4mt).
\end{equation}
 
The spin Josephson effect can be probed by trying to detect 
the {\it induced} electric fields associated with it 
\cite{Loss}. It can be seen from Maxwell equations
that a spin current should induce an electric field 
in the same way that a charge current induces a magnetic 
field. Indeed, in absence of voltage the induced electric 
field should satisfy  

\begin{equation} 
\partial_xE_y-\partial_yE_x=-4\pi\mu_B\partial\langle S_z\rangle/\partial t
=4\pi I_J^{\rm spin},
\end{equation} 
where we have assumed that $\partial{\bf H}/\partial t=0$. 

In conclusion, we have shown that a spin Josephson effect should 
occur in FM/FM tunnel juntions. In the more general case where 
the FM order parameter is three-dimensional, the effect is necessarily 
an AC-like one, with the oscillations in time being associated to the 
$z$-component of the magnetization. Finally, we would like 
to stress that our analysis is completely different from  
Josephson effects discussed previously in FM/FM junctions where 
a {\it charge} current is discussed \cite{Su}. Here we have 
considered the phase coherence of a {\it spin} tunnel current. 
Our point of view is similar to the one of Ref. \cite{Konig} 
where averages like $\langle c_\uparrow^\dagger c_\downarrow\rangle$ 
also play a crucial role. In Ref. \cite{Konig} the 
phase coherence arises from the fact that a thin film helimagnet is 
being considered while in our case it is due to the existence 
of a boundary between two ferromagnetic metals.

\acknowledgments  
F.S.N. was supported by DFG Sonderforschungsbereich 290.


\begin{thebibliography}{100}

\bibitem{Spintronics} \Name{Prinz, G.A.} 
\REVIEW{Science}{282}{1998}{1660}.

\bibitem{Hirsch} \Name{Hirsch, J.E.} 
\REVIEW{Phys. Rev. Lett.}{83}{1999}{1834}.

\bibitem{Brataas} \Name{Brataas, A.,  
Tserkovnyak, Y. \and Bauer, G.E.W. \and 
Halperin, B.I.}
\REVIEW{Phys. Rev. B}{66}{2002}{060404(R)}.

\bibitem{Wang} \Name{Wang, B., Wang, J. \and Guo, H.}, 
Phys. Rev. B {\bf 67}, 092408 (2003).

\bibitem{Sloncz} \Name{Slonczewski, J.C.} \REVIEW{J. Magn. Magn. 
Mater.}{126}{1993}{374}.

\bibitem{Josephson} \Name{Josephson, B.D.} 
\REVIEW{Phys. Lett.}{1}{1962}{251}.

\bibitem{Su} \Name{Su, G. \and Suzuki, M.} 
\REVIEW{Mod. Phys. Lett. B}{16}{2002}{711}; 
for an earlier closely related analysis, see  
\Name{Berger, L.} 
\REVIEW{Phys. Rev. B}{33}{1986}{1572}.

\bibitem{Borovik} \Name{Borovik-Romanov, A.S.,  Bunkov, Yu.M.,  
Dmitriev, V.V., Mukharskiy, Yu.M. \and Sergatskov, D.A.} 
\REVIEW{Phys. Rev. Lett.}{62}{1989}{1631}; 
\Name{Borovik-Romanov, A.S.,  Bunkov, Yu.M.,   
de Vaard, A.,  Dmitriev, V.V.,  Makrotsieva, V.,   
Mukharskiy, Yu.M. \and Sergatskov, D.A.}
\REVIEW{JETP Lett.}{47}{1988}{478}.

\bibitem{Fomin} For a review see \Name{Fomin, I.A.}
\REVIEW{Physica B}{169}{1991}{153}. 

\bibitem{Konig} \Name{K\"onig, J., B{\o}nsager, M.Chr. \and 
MacDonald, A.H.} 
\REVIEW{Phys. Rev. Lett.}{87}{2001}{187202}; 
\Name{Heurich, J., K\"onig, J. \and MacDonald, A.H.}
\REVIEW{Phys. Rev. B}{68}{2003}{064406}.

   
\bibitem{Wen} \Name{Wen, X.G. \and Zee, A.} 
\REVIEW{Phys. Rev. Lett.}{69}{1992}{1811};  
\REVIEW{Phys. Rev. B}{47}{1993}{2265}.

\bibitem{Ezawa} \Name{Ezawa, Z.F. \and Iwazaki, A.}
\REVIEW{Phys. Rev. B}{48}{1993}{15189}.

\bibitem{Stern} \Name{Stern, A., Girvin, S.M., MacDonald, A.H. \and Ma, N.} 
\REVIEW{Phys. Rev. Lett.}{86}{2001}{1829}.

\bibitem{Balents} \Name{Balents, L. \and Radzihovsky, L.} 
\REVIEW{Phys. Rev. Lett.}{86}{2001}{1825}.   

\bibitem{Spielman} \Name{Spielman, I.B., Eisenstein, J.P., Pfeiffer, L.N.  
\and West, K.W.}
\REVIEW{Phys. Rev. Lett.}{84}{2000}{5808}.

\bibitem{Lozovik} \Name{Lozovik, Y.E. \and Poushnov, A.V.} 
\REVIEW{Phys. Lett. A}{228}{1997}{399}. 

\bibitem{Fogler} \Name{Fogler, M.M. \and Wilczek, F.}
\REVIEW{Phys. Rev. Lett.}{86}{2001}{1833}.

\bibitem{Anderson} \Name{Anderson, P.W.}
\Book{Lectures on the Many-Body Problem} 
\Editor{E.R. Caianello}
\Publ{Academic Press, New York}
\Year{1964}.

\bibitem{Note1} The main feature of systems 
exhibiting Josephson effects is that the phase is canonically 
conjugate to an operator whose time evolution gives the current. 
There are many examples of Josephson-like  
effects in physical systems other than superconductors. 
Josephson effect occurs also in Bose-Einstein consensed 
systems [\Name{Cataliotti, F.S., Burger, S., Fort, C., Maddaloni, P.,  
Minardi, F., Trombettoni, A., Smerzi, A. \and Inguscio, M.} 
\REVIEW{Science}{293}{2001}{843}]  
and in superfluids $^3$He 
[\Name{Davis, J.C. \and Packard, R.E.} 
\REVIEW{Rev. Mod. Phys.}{74}{2002}{741}] and $^4$He 
[\Name{Sukhatme, K., Mukharsky, Yu., Chui, T. \and 
Pearson, P.}
\REVIEW{Nature}{411}{2001}{280}]. 




\bibitem{Villain} \Name{Villain, J.}
\REVIEW{J. Phys. (Paris)}{35}{1974}{27}.



\bibitem{Stoner} \Name{Stoner, E.C.}
\REVIEW{Proc. R. Soc. London A}{154}{1936}{656}.

\bibitem{Tinkham} \Name{Tinkham, M.} \Book{Introduction to 
Superconductivity, 2nd Ed.} \Publ{McGraw-Hill} \Year{1996}.

\bibitem{Loss} \Name{Loss, D. \and Goldbart} \REVIEW{Phys. Lett. A}{215}
{1996}{197}; \Name{Meier, F. \and Loss, D.} \REVIEW{Phys. Rev. Lett.}
{90}{2003}{017205}; \Name{Sch\"utz, F., Kollar, M., \and 
Kopietz, P.} \REVIEW{Phys. Rev. Lett.}{91}{2003}{017205}.


\end{thebibliography}
\end{document}